\documentclass[letterpaper]{article} 
\usepackage{aaai24}  
\usepackage{times}  
\usepackage{helvet}  
\usepackage{courier}  
\usepackage[hyphens]{url}  
\usepackage{graphicx} 
\usepackage{natbib}  
\usepackage{caption} 
\usepackage{algorithm}
\usepackage{algorithmic}

\usepackage{newfloat}
\usepackage{listings}
\DeclareCaptionStyle{ruled}{labelfont=normalfont,labelsep=colon,strut=off} 
\floatstyle{ruled}
\newfloat{listing}{tb}{lst}{}
\floatname{listing}{Listing}
\usepackage{tabularx}
\usepackage{rotating}
\usepackage{multirow}
\usepackage{color}

\title{Understanding the Impact of Physicians' Legal Considerations on XAI Systems}
\author{
    Gennie Mansi,
    Mark Riedl
}
\affiliations{
    \textsuperscript{\rm 1}Georgia Institute of Technology\\

    Atlanta, GA\\
    gennie.mansi@gatech.edu,
    riedl@cc.gatech.edu
}

\begin{document}

\maketitle

\begin{abstract}
Physicians are---and feel---ethically, professionally, and legally responsible for patient outcomes, buffering patients from harmful AI determinations from medical AI systems. Many have called for explainable AI (XAI) systems to help physicians incorporate medical AI recommendations into their workflows in a way that reduces the potential of harms to patients. 
While prior work has demonstrated how physicians' legal concerns impact their medical decision making, little work has explored how XAI systems should be designed in light of these concerns. In this study, we conducted interviews with 10 physicians to understand where and how they anticipate errors that may occur with a medical AI system and how these anticipated errors connect to their legal concerns. In our study, physicians anticipated risks associated with using an AI system for patient care, 
but voiced unknowns around how their legal risk mitigation strategies may change given a new technical system. Based on these findings, we describe the implications for designing XAI systems that can address physicians' legal concerns. 
Specifically, we identify the need to provide AI recommendations alongside contextual information that guides their risk mitigation strategies, including how non-legally related aspects of their systems, such as medical documentation and auditing requests, might be incorporated into a legal case.

\end{abstract}

\section{Introduction}

Medical AI has been championed as a path to reducing barriers to access and increasing quality of care for patients, including   ``optimizing workflows in hospitals, providing more accurate diagnoses, and bringing better medical treatments to patients'' \cite{GerkeEtAl2020}. Many have looked to explainable AI (XAI) to help physicians and other healthcare workers incorporate AI systems into their workflows. Explanations are intended to help physicians and healthcare workers act in response to AI systems, preventing or mitigating negative outcomes 
such as incorrect treatments~\cite{AIAAIC_Watson}\footnote{The AIAAIC is regularly cited and mentioned by academic, non-profit, advocacy, government, and commercial research institutes, associations, and agencies \cite{AIAAIC_Mentions}; https://www.aiaaic.org/};
misdiagnoses of heart attacks~\cite{AIAAIC_HeartAttacks, TechCrunch_HeartAttacks}; and
the accentuation of historical and systemic bias for marginalized groups~\cite{Benjamin2019, AbramoffEtAl2020}. 

To support physicians, AI explanations need to help them address their contextual needs.
Physicians are tasked with making decisions at the intersection of medical knowledge and patients' values and individual needs \cite{MolemanEtAl2021}. Physicians are---and feel---accountable for the well-being of their patients and their own livelihoods as they use medical AI systems. 
For example, physicians must often make nuanced decisions, such as helping patients decide between cancer treatments to support their quality of life or administering a medication for an unapproved use when other treatments are not working, which may not be recommended by an AI system. Physicians must also advocate for patients, sometimes going against AI recommendations that may be incorrect, or against insurance companies that unjustly deny coverage to patients \cite{Statnews_Medicare}. 

In addition to their medical and relational needs, physicians are also impacted by legal considerations. In high-stakes decision making environments such as medicine, laws and regulations are an important infrastructure that shapes users' context \cite{MansiEtAl2025_LegalXAI}. Physicians have been known to change their behavior in response to perceived legal risks~\cite{VelthovenAndWijck2012}, including the kinds of tests they order and how frequently they refer patients to specialists \cite{CarrierEtAl2010}.
Physicians' risk mitigation strategies center around anticipating potential sources of errors for which they could be held legally liable. For example, ordering extra tests and referring patients to specialists can provide extra legal buffer against claims of negligence in a malpractice suit \cite{CarrierEtAl2010}, which can be brought against a physician when an error occurs that significantly harms a patient. 

Despite its significant impact on physician behavior, little work has explored the legal dimensions of physicians' information needs from an AI system, or how they may adapt their decision making with the system to protect themselves against perceived harms from errors. 

In our study, we seek to understand how they connect anticipated errors with legal risks, their subsequent information needs from AI explanations, and the corresponding actions medical AI needs to enable. We conducted interviews with medical physicians to answer the following research questions:
\begin{itemize}
    \item \textbf{What do physicians' perceive as sources of error when using AI systems?}
    \item \textbf{How do physicians' perceptions of potential errors relate to their perceptions of legal risk when using AI systems?}
\end{itemize}

We analyze the transcripts of semi-structured interviews with 10 physicians to understand how laws and regulations shape physicians' context and the design implications for AI explanations. We highlight how physicians primarily anticipate errors around how the AI system is designed and how physicians will use the AI system---both of which they face in their current uses of largely non-AI based systems. Participants highlighted that their current primary risk mitigation strategy, using their clinical judgment, was still the best way to protect patients and minimize risk to themselves and others. However, they also noted several shifts or unknowns in how their clinical judgment and other legal risk mitigation strategies would change with an AI system. With respect to medical judgment, they worried about their own (and others') potential to default to an AI decision given the fast-paced environment of many hospitals. They also described how their defensive medical practices that control for legal risk might change, including impacts to medical documentation practices and how to identify and notify others of potential errors. Based on these concerns, we highlight several design implications to build XAI systems that can better address physicians' legal needs, enabling them to better protect patients and themselves.

\section{Background and Related Work}

\subsection{Understanding the Medical-Legal Context}

Lawyers create laws to directly shape the design and deployment of AI systems to protect consumers. AI is what lawyers call a ``credence good''---a consumer can't know the quality of the good until after the purchase. It requires blind faith in the quality or some rigorous, systemic evaluation in order to ensure it works properly \cite{Price2022}. Public access to generative AI has magnified existing conversations among legal scholars about regulating AI-based technologies in medical settings. While many are optimistic about the potential for AI to improve patient care \cite{Economist_MedicalAI}, there are significant concerns around how AI algorithms may actually increase bias and harms, such as biasing the information provided to the human monitor, creating the potential for amplified harms \cite{AbramoffEtAl2020}.

Due to the risk consumers assume with AI as a credence good, lawyers regulate its development at multiple points through \textit{ex ante} and \textit{ex post} laws. 
\textit{Ex ante} laws  impact technology design before it enters stakeholders' workflows. For example, current FDA regulations ensure and validate the safety of technology before physicians' use.  \textit{Ex post} laws  define a procedures for determining and mitigating harm after a technology is distributed to consumers. Medical malpractice laws that determine fault when patients are harmed are examples of \textit{ex post} regulations. Further, \textit{ex ante} and \textit{ex post} regulations can interact with each other. Under previous FDA regulations, achieving Class III FDA approval of a medical technology could shield a company from certain degrees of liability claims if a consumer was harmed \cite{FDA_Liability}. In this way, \textit{ex ante} and \textit{ex post} laws change the AI value chain from design and development to deployment. When creating a new AI-based technology, developers and companies must consider both \textit{ex ante} and \textit{ex post} laws in their designs and verification processes.

\subsection{Understanding Physicians' Legal Concerns}
Among medical physicians, concerns around malpractice liability are pervasive and have been shown to affect the way physicians provide care to patients \cite{VelthovenAndWijck2012}. These concerns cause physicians to adopt ``defensive medicine'' changing their clinical practice to address perceived risk \cite{NashEtAl2004}. This can take the form of ``positive defensive medicine'', such as ordering more tests and procedures and taking more time to explain risks \cite{CarrierEtAl2010, Dickens1991, NashEtAl2004}; or ``negative defensive medicine'', such as restricting the the scope of their practice and instead referring patients to other specialists \cite{CarrierEtAl2010, Dickens1991, NashEtAl2004}. The effects of such pressure on physicians has mixed results---it's not clear that physicians' concerns about malpractice improve the quality of care \cite{VelthovenAndWijck2012}, even though it raises the cost of health care \cite{CarrierEtAl2010}. Further, both the threat of and actual legal process have been associated with psychological, physical, and behavioral practice changes, including depression and adjustment disorder \cite{NashEtAl2004}.

Further, research has shown that physicians don't accurately understand their legal risk despite changing their clinical practice to reduce it \cite{CarrierEtAl2010, ZajdelEtAl2013}. Several studies have evaluated that changes in policy don't actually impact clinicians perception of legal risk \cite{CarrierEtAl2010, Dickens1991}. This indicates that while actual legal risk matters to the development of AI explanations, physicians' perceptions of legal risk holds significant implications for the kind of explanations offered by AI systems.

Second, the medical community has foregrounded the use of clinical judgment as key to enabling proper human oversight over medical AI \cite{NyrupAndRobinson2022, AbramoffEtAl2020, FroomkinEtAl2019}. Because both AI systems and physicians could be considered ``experts of sorts'' \cite{Chan2021}, it is unclear how to design AI systems that provide balance with clinical judgment to improve overall patient care. One group of researchers advocated for ``unremarkable AI,'' that is, AI integrated directly into clinical decision making contexts while remaining unintrusive to overall medical practices and norms \cite{YangEtAl2019}. For example, these authors embedded AI decision models on the corner of PowerPoint slides used in discussions around complex patient care, so physicians and other medical workers could reference it as a part of their normal work practices. Others have promoted the need for \textit{trust calibration}, supporting users in knowing when to trust an AI decision. \citet{ZhangEtAl2020} discussed providing the right information in AI explanations to align the ``error boundaries'' of AI systems and users, such that the decisions in which an AI system is most likely to make a mistake are complemented by human knowledge and vice versa. 

In light of this information, legal information about an AI system can affect how physicians exercise their clinical judgment. If physicians believe they will be liable if harm comes to a patient when they disregard an AI system's recommendation, then they may be hesitant to exercise their clinical judgment. If they believe they are ultimately responsible for harm to the patient, regardless of how they use the AI system, then they may choose to disregard the AI system altogether, and rely only on their clinical judgment, nullifying the potential benefits of the AI's recommendation.

Consequently, there are at least two dimensions on which understanding needs to be calibrated: (a)~mitigating personal liability risk and medical decision-making, and (b)~AI systems need to be designed to support physicians actions as they consider both. The HCI community has recognized that XAI solutions that address explainability needs should not be limited to algorithmic explanations or showing model internals \cite{SunEtAl2022}. As a result, AI systems should not only include algorithmic explanations but also address physicians' information needs around legal liability.

\subsection{Expectations around Explainable AI}
Because legal regulations impact the design of software, it's important to enable communication between professionals in law and computer science \cite{MansiEtAl2024_Contestability}. Professionals in computer science and law have both expressed the desire for transparency and accountability in AI systems deployed in high stakes applications, and they have focused on Explainable AI systems to meet this need. Among computer scientists, XAI refers to AI systems that provide human-understandable explanations for their reasoning or responses~\cite{Lipton2018, ChariEtAl2020, GuidottiEtAl2018}, and it has a core goal of enabling action. On the other hand, lawyers highly prize transparency, and they desire Explainable systems as a path to looking ``under the hood'' of complex systems, preventing injury or determining fault in the case of harm \cite{BaitAndSwitch2023, FroomkinEtAl2019, DesaiAndKroll2017}.

While both lawyers and computer scientists share the call for XAI, they are also both struggling with and occasionally conflict in with how they see AI systems providing explanations that enable desired user actions.  
Historically, computer scientists have invested primarily in the algorithmic advancement of XAI systems \cite{AbdulEtAl2018}, but recently, there's a greater call to consider users' environment \cite{AbdulEtAl2018} with many human-computer interaction (HCI) researchers drawing attention to the gap in delivering satisfying user experiences \cite{LiaoEtAl2020}. There's also contention among lawyers as to the role of XAI, and some have expressed frustration or distrust with the computer science community around prevailing explanation methods. For example, Babic and Cohen \cite{BaitAndSwitch2023} criticize post hoc explainability methods as ``insincere by design'', rendering them of little value to legislators who they feel are right to feel transparency.

Further, in order to ensure beneficial policies and the effective deployment of AI systems, both the legal and computer science communities need to understand users and their needs. Effective communication and understanding around users' needs can help guide the creation of laws that have their intended effect on technology design and use; it also helps AI Creators identify and implement technical methods for users and lawyers to more easily pursue avenues for change when adverse outcomes happen \cite{MansiEtAl2024_Contestability}. Consequently, in our study we describe physicians' legal concerns as a first step towards supporting a broader understanding of the context in which medical AI systems may be used and how that connects to users' current challenges. By beginning where the users are---that is, what physicians perceive as legal threats---lawyers and technologists can begin to create a shared understanding both of users and of what XAI systems need to include.

\section{Methodology}
We used qualitative design methods to understand the errors physicians anticipate from the use of AI systems during patient care and how they may adapt their risk mitigation strategies in response. %
We used scenario descriptions around tasks familiar to physicians to prompt them to think deeply about their approach to their work environments and how different aspects of that situation impact their effectiveness to \textit{act} with other tools, people, and their own knowledge.

\begin{figure*}[t]
  \includegraphics[width= 1\linewidth]{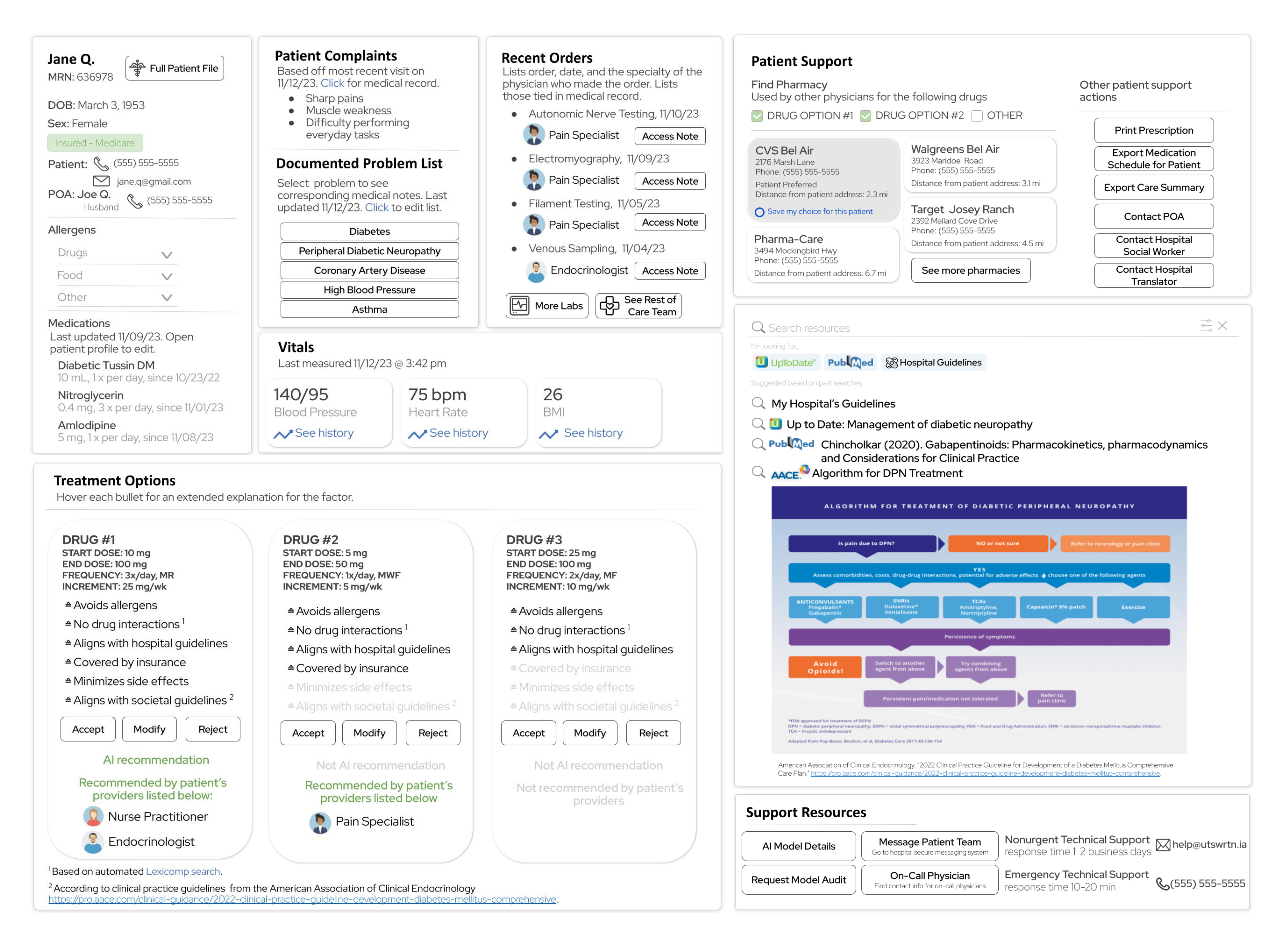}
  \caption{\textbf{Visual Used in Interviews}. The visual presented to physicians in the study. The visual represented an AI system based on an electronic health record system. It was intended to prompt participants to think deeply about their approach to their work environments and how different aspects of that situation impact their effectiveness to act with other tools, people, and their own knowledge. }
 \label{fig:MedicalInterface-small}
\end{figure*}

To ease cognitive load, we created and used a visual of a hypothetical medical AI system that was designed as an imaginary electronic-health record-like system
(see Figure~\ref{fig:MedicalInterface-small}). 
We designed the visual to align with active efforts by commercial developers  \cite{DOSIS, Pfizer_MedicalAI, FDA_AIMedical, Zawacki-RichterEtAl2019} and reflect the many tools already familiar to participants. 
The visual was not intended to be a real practical system, but instead could provide participants a familiar place to jump off and generate a broader set of ideas given the scenario.
The visual also went through two rounds of iterative design and piloting with physicians to ensure its fidelity to their workplaces and readability. 

In our interviews, we asked participants to place themselves in the role of an attending physician at a large hospital who decides on drug dosages for a patient with the help of other healthcare professionals. The participants had to make a decision about a prescription for a patient using an AI-based recommendation system. We showed them the visual and asked them to describe the kinds of tasks they might need to accomplish, people they might need to talk to, and goals they might have. Participants were instructed to rely on their past and present experiences when explaining the types of information they would use, the way they would use them, and the reasons for those choices. After situating physicians' in the use case scenario, we then asked four questions to elicit physicians' perceptions of how such a system might fit into their broader decision making and environment:
\begin{enumerate}
    \item What are the ways in which you would feel confident about using a similar AI interface to care for a patient? Is there different or additional information that you would need to feel confident?
    \item What are the ways in which you would not feel confident about using a similar AI interface? Is there additional information that could help you feel confident?
    \item What kinds of mistakes do you think the computer might make?
    \item What kinds of mistakes do you think you and other physicians might make as they work with the computer?
\end{enumerate}

We intentionally phrased these questions broadly to avoid explicitly pinpointing legal risks and language around ``errors'' to avoid priming participants towards particular responses. Thus we also aimed to encourage them to describe a variety of and interactions between medical, relational, legal, and other overlapping needs from explanations.

\subsection{Recruitment}
We recruited physicians who would be direct users of the AI systems in the scenarios with a range of time-in-service and expertise. The recruitment was initiated through online messaging and contacts of the research team, followed by snowball sampling. For physicians, the recruiting criteria were: must have an MD and at least be in their medical residency. We verified these criteria were met through correspondence with participants prior to interview. Informed consent was obtained from all participants prior to participation.

A total of 10 physicians (labeled V1-V10) were recruited. Physicians ranged in their expertise from second year residents to experienced clinicians with research experience. Both V5 and V8 had extensive experience teaching junior physicians and peers. V9 had a PhD in research with a clinical focus. Of the medical residents, they ranged from in experience from second to fellowship years, with the majority as third or fourth year residents. 

Our needs gathering process took a depth rather than breadth approach, resulting in relatively small sample size of physicians for our interviews. However, we reached saturation as physicians began to repeat information. Data from scenario descriptions depend on the quality scenario created, the strength of the connection between participants and the scenario, and the variability of participants' experience. While we address these limitations through recruiting and the careful, iterative design of the study probe, insights should be viewed as formative.

\subsection{Analysis}
The semi-structured interviews were conducted online with screen-sharing for the interviewee and lasted 58 minutes on average. All interviews were video-recorded to include the interface activities, then transcribed for analysis. We used a combination of grounded theory \cite{StraussAndCorbin1995} and thematic qualitative analysis \cite{MilesEtAl2020} to analyze our data for the errors participants' anticipated and how they would respond to the XAI system and adapt their risk mitigation strategies.

\section{Findings}

As discussed in the Related Works section, physicians' risk mitigation strategies center around anticipating potential sources of errors for which they could be held legally liable. When an error occurs that significantly harms a patient, patients can make a claim of negligence in a malpractice suit against a physician.  Because physicians react to perceived legal risks (which may not be reflective of actual legal risks), it's important to understand both what kinds of errors physicians anticipate when working with AI and risks they explicitly label as a legal risk.

\subsection{Perceived Risks with AI Systems}
Physicians identified potential risks across two broad areas: (1)~those presented by the way an AI system is designed, and 
(2)~those presented by the way they and other physicians would use the system. 

\subsubsection{Risks Presented by System Design} Participants were concerned about errors in the way the system was implemented, such as those caused by AI Creators' improper expertise or logic (V2, V5, V8), and those caused by fundamentally incorrect assumption about appropriate applications of AI systems (V8). 

Participants also expressed concerns around risks to patients presented by the data used to create the system. Physicians voiced concerns about 4 sources of potential errors. First, they identified the potential for errors caused because of the \textbf{inability of data to reflect or capture critical clinical interactions}, including those presented by past clinical experience (V1, V6, V7), bedside interactions (V1, V6, V7), social or psychological issues (V5, V6, V7), specific patient differences (V1, V6, V9), and physical examinations (V1, V6, V7). Second, they were concerned about the \textbf{representativeness of the data}, how it may not reflect patient diversity (V9) and location-specific factors (V2, V9), and how a poor AI policy may become a ``self-fulfilling prophecy'' in which poor recommendations are reinforced by outputting similar results, amplifying trends in the data (V9).  Third, physicians worried about how recommendations from the \textbf{AI system may differ from societal guidelines}, including adapting to guidelines as they are introduced or updated (V4, V7, V9). Finally, physicians were concerned about how  an AI system trained with or relying on the data in an electronic health record (EHR) systems may lead to erroneous outputs simply because of the \textbf{potential inaccuracies in the data}, such as those commonly caused by human errors when inputting data into a medical record.

\begin{figure}[h!]
  \includegraphics[width=0.47\textwidth]{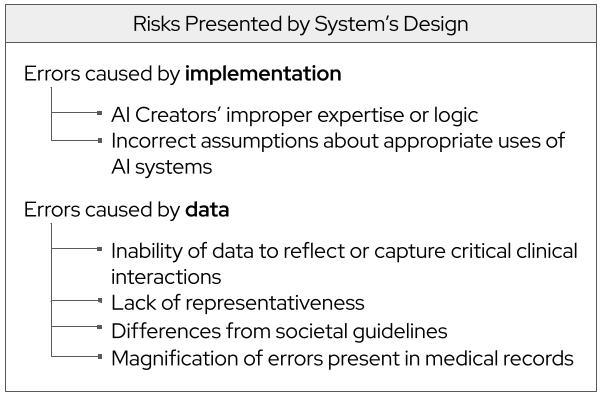}
  \caption{The risks that participants identified around the system design mainly focused on how the system was implemented and what kind of data was used to train the system.}
 \label{fig:Risks_SystemDesign}
\end{figure}

\subsubsection{Risks Presented by Physicians' Use of the System} Participants separated the risks in physicians' use of the system into two periods: (1) the period when physicians are calibrating to the use of a new system, covering the mistakes made in adjusting to decision-making based on a current understanding of the system's strengths and weaknesses, and (2) the period of continued use, covering the persistent issues that continue throughout deployment. 

By far, participants described more potential risks as physicians calibrate to the system than risks as they continue to use the system. Participants described the need for physicians to build an appropriate (dis)trust of the system based on $\bullet$~cross references (e.g. PubMed, Up-to-Date) that are created independently of the system (V2, V3, V4, V7, V8, V9), $\bullet$~the qualifications of those who built and deployed the system (V3, V6), and $\bullet$~independent medical practice norms (V2, V6, V8). They described how risks are presented by the physicians inability to anticipate and predict the consequences around $\bullet$~how the system will deal with errors due to messy digital data (V6); $\bullet$~how to adapt their understanding, decision-making, and actions given the evolving nature of the technology (V3, V4, V7, V8, V9, V10); $\bullet$ how to rely on others' uses of the recommendations, including reporting errors (V4, V5, V7); $\bullet$~and how to calibrate to the ways the system does (not) meet their standard of care or logic (V3, V6, V8, V9). Significantly, physicians worried about how they should document poor system performance or patient outcomes in order to help fix errors with the system (V5, V8). They also were concerned about the negative impacts to patients from trial and error as physicians learned to rely on the system (V5, V6, V8).

\begin{figure}[h!]
  \includegraphics[width=0.47\textwidth]{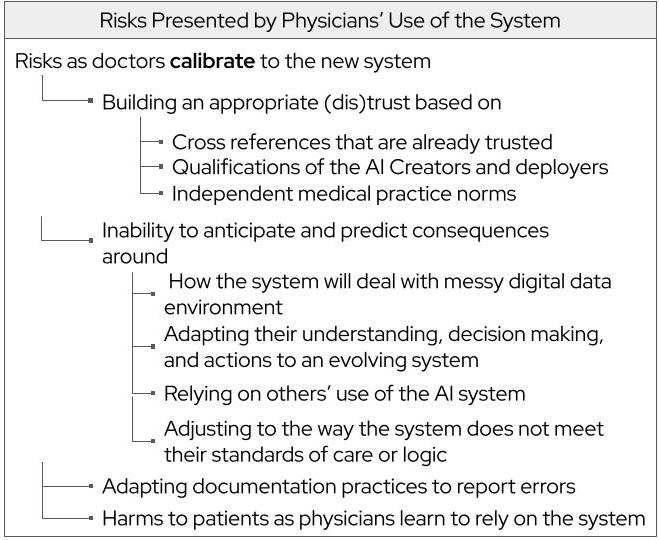}
  \caption{Participants associated the most risk and potential for errors as physicians calibrated to using a new system.}
 \label{fig:Risks_Calibrating}
\end{figure}

The primary risk that participants anticipated persisting throughout deployment was that physicians would not exercise the appropriate medical expertise and skills they were trained with. Examples included: $\bullet$~narrowing thinking too about potential causes or treatments (V3); $\bullet$~overlooking key details in a patients medical file (V2, V4); $\bullet$~not applying their own expertise to play a role in decision making (V1, V2, V3, V4, V5, V8); $\bullet$~and defaulting to the AI recommendation automatically (V3-V10).

\begin{figure}[h!]
  \includegraphics[width=0.47\textwidth]{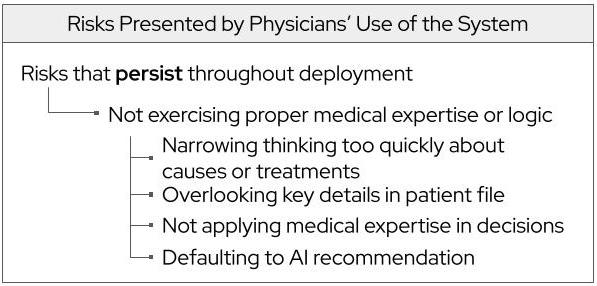}
  \caption{The primary risk that participants anticipated persisting throughout deployment was that physicians would not exercise the appropriate medical expertise and skills while using the AI system and its recommendation.}
 \label{fig:Risks_Persistent}
\end{figure}

\subsection{Physicians' Legal Concerns around AI Systems}

Our analysis indicates how the law currently impacts how physicians think about treating patients. Every physician (V1-V10) stated concerns around physicians' over reliance on the AI system without engaging in critical thinking about the diagnosis, something that could constitute negligence and be grounds for a malpractice claim. For example, V3 said ``So I really think like the physician's brain has narrowed into like, one thing the computer tells it, and might just disregard... a different scenario. If the patient complaint was heart like chest pain, and this paradigm told me it was reflux. And I just... do the reflux thing and miss the heart attack. Like, we're in deep trouble.'' 
Likewise, V9 indicated this might be especially easy to slip into with tight time constraints, ``I think when it gets like really busy, it's like very tempting to just like, not have to think about things anymore.''

Further, several physicians explained defensive medical practices that they explicitly tied to controlling legal risk. For example, V7 mentioned that adhering to institutional guidelines ``protects you from liability.'' 
Both V7 and V6 also mentioned defensive medical practices around asking physicians in other specialties before prescribing medications. For example, V7 said:
\begin{quote}
``There's a component of liability to all of this, right?...I'm the prescribing physician, it's still 100\% on me. If I have a conversation with the patient's endocrinologist, or document that I reviewed documentation from a specialist, that makes me much more comfortable prescribing the medication.''
\end{quote}
V6 expressed a similar hesitancy, explaining how they were unsure of the way an AI system might change this practice: ``So I think it [consulting someone from that speciality] has to be documented in the system. I don't have a great plan for how this [documenting decisions around an AI recommendation] should work out, because it has kind of a legal part to it.'' 

Not only did participants mention perception around legal risk, but they also expressed uncertainty around how an AI-based system would impact their risk mitigation strategies. Three of the ten physicians (V5-V7) echoed concerns around how an AI-based system may change current norms used to mitigate legal risk. For example, V5 had concerns about how documenting their acceptance or rejection of the AI recommendation could impact their legal liability,
\begin{quote} 
``Well, if there's a lawsuit, I don't know what the legal implications would be. Someone could say to me, Oh, you rejected this because it didn't match hospital guidelines, or societal guidelines, but which ones are referring to? How well do you know these guidelines? I guess I shouldn't be thinking legally... [But] you [are] taking responsibility for knowing, knowing that those have been considered in this patient and considered to be optimal or positive for the patient's benefit. So I guess I'm a little concerned about how this data would be used.''
\end{quote}
V8 expressed how their understanding of malpractice informed a more critical evaluation of the skills of the person auditing the AI system,

\begin{quote}
``Because it's malpractice [not to follow the correct guidelines]...so if you're questioning the AI, and you want an audit, then you want somebody who's more knowledgeable of the literature and smarter in that area than you are [to review the AI]. So if you're having questions, you'd like somebody with knowledge to answer them...you would have to be an expert in some field.''
\end{quote}

\subsection{Clinical Medical Decision Making: Decreasing Risk, Increasing Confidence}

Physicians viewed their critical thinking as the path to reducing risk, and they described treating the AI more like a ``thinking buddy'' rather than a solution finder. Across all risks identified, whether they were risks presented by system design or those presented by other physicians', our participants consistently highlighted the importance of relying on clinical decision making to mitigating risks. In fact, participants tied all risks that were persistent throughout deployment to the failure to appropriately exercise their decision making. 

The order in which participants narrated how they would navigate all the information in the interface reflected this emphasis on clinical decision making. All but one physician (V1-V9)
explained that they would begin with the patient summary information, check their thinking with a medical algorithm, and then look to the AI recommendation as a final check before deciding on a treatment. Even V10, who described a different order and approach to the visual, expressed the idea that they were ``already coming up with like, something in the back of my mind that like I would want to do and then I think I would try to verify with the AI.'' Additionally, V4 explained how this process minimized the risk in using the AI, ``I think that it's [using the AI is] pretty low stakes, as long as you are still using your brain and your clinical judgment. And so I think if you're a provider who takes great care, and like your clinical judgment, and things like that, I don't think there's a ton of hesitancy in using it. Because the worst comes to worst is they give you three options. And you're like, those are horrible options. I don't agree with that, and you do your own thing, and then there's no harm no foul.''

\section{Discussion}
In a survey of US physicians, 78\% of participants agreed strongly that it is becoming increasing risky to rely on clinical judgment, rather than diagnostic testing \cite{CarrierEtAl2010}. Further, this concern around malpractice fear has been shown to have a stronger impact on physicians with fewer years experience, as opposed to those with more experience~\cite{CarrierEtAl2010, CatinoAndCelotti2009}. This is further complicated by the introduction of an AI system, which could be seen as another diagnostic measure against which physicians are held accountable. Our interviews also indicate that AI-based decision support can add further nuance to the tension between clinical decision-making and diagnostic testing. As one of our participants stated, ``As a trainee, I want to be developing my own clinical thinking. And so I'd be worried that relying on AI would just turn me into like a non skilled worker and get in the way of developing those skills'' (V9). 

The narrow road to decreasing risk and increasing confidence of using AI systems for medical care is reflected in the uses of medical AI systems for which participants felt more comfortable. Participants felt more comfortable with AI systems that optimized drug doses (V5); supported their own or others' education (V4, V9); uncovered new, updated, or unknown guidelines (V1, V2, V8); and helped them expose blind blind spots in their medical decision making, such as helping them consider other treatment options or improving their decision making heuristics (V1, V4, V5). However, participants' positive hopes around what AI could afford in terms of benefits were in tension with their experiences with their current Electronic Health Record (EHR) %
systems and the potential for significant, negative outcomes. For each of the benefits mentioned, participants identified downsides or concerns around why they may not reap the benefits. Participants contrasted the positive benefits of medical AI systems with their own potential to misuse them by defaulting to their recommendations without properly applying their clinical decision making skills, especially in light of the significant time constraints 

Participants also described legal concerns around their medical practice and changing their behavior when using an AI system in order to mitigate perceived legal risks. They described ways that they would continue certain risk mitigation practices, such as following institutional or societal guidelines (V2, V7, V9) and consulting or referencing other providers' opinions of their patients (V3, V4, V7, V8). 
Importantly, participants also discussed how their current legal risk mitigation strategies were thrown into question with the introduction of AI systems.  Participants questioned if and how they should adapt their medical documentation practices (V3, V5, V8), and change decision making and report errors in response to errors in system outputs (V3, V4, V7, V9) in light of new potential risks presented by the system. 

Participants highlighted how medical AI systems do not just change how they mitigate risks in their own clinical decision making---they also can change how they see or mitigate risks with other providers. For example, V7 voiced concerns about the use of AI systems by non-physician providers such as nurse practitioners and physician assistant, who don't receive the same level of training and may not notice or pursue the same kinds of errors as a medical physician. V4 noted the questions around if and how the head of a medical practice should rely on those they employ to engage in the same level of rigor in clinical decision making as themselves, ``Can I trust every person to still use their brain, to use this to supplement your years of experience and your education?'' Because physicians workflows are naturally shaped around ways of mitigating liability, AI systems don't only disrupt their own clinical decision making, but also disrupt how they think about their responsibilities and relationships with other healthcare workers.

\subsection{Toward Designing XAI Systems to Meet Physicians' Legal Needs}

AI is most helpful in the case when a physician is unsure of what to do or when they might accidentally be missing an aspect of patient care. However, most of the physicians described completely disregarding the AI if it disagreed from what they would do because it might open them to additional liability. This could lead to potential confirmation bias and the nullification of positive effects of AI as physicians try to mitigate their legal risk. 
Some researchers have stated the need to address ``physicians' fear of litigation'' and to remove the ``fear of `failure to care' '' \cite{NashEtAl2004}. However, doing so with AI systems entails clearly communicating legal risks and understanding physicians' corresponding information needs and actions. With a new generation of physicians ``growing up'' with these technologies,
it's important to treat the law as a contextual factor that is taken into account when designing AI explanations. 

Our findings point to some initial implications for how we can begin to further understand and adapt AI explanations to meet physicians' legal considerations.

\paragraph{A Physician-Centered Understanding of XAI}

Our participants fluidly considered information from the visual, their scenario, and their clinical experiences to predict and understand potential medical and legal risks to themselves and their patients. Participants referred to the entire system as the ``AI system'', and they did not distinguish between the AI, XAI, and interface components. Instead, they evaluated all information presented to them as a whole to understand and predict risk. Significantly, this also meant that if one aspect of the system did not align with their standards or logic, they would distrust the rest of the system. As V7 explained, ``if there were a discrepancy between like this [AI recommendation], and then what I've reliably been using, I would definitely defer to what I've been reliably using... [If a known risk of a medication] wasn't listed [in the AI reasoning], [it's a] huge red flag, it would make me worried, like, about everything else being trustworthy.'' This included information in the visual that we explicitly told participants was not AI-generated or produced. 

Similarly, participants interpreted the AI recommendation alongside other trusted resources referenced in the visual, such as other physicians' notes, and external references that are currently a part of their workflows (e.g., UpToDate\footnote{A well-trusted clinical reference often used by physicians to access medical information and recommendations; https://www.wolterskluwer.com/en/solutions/uptodate} or PubMed). While the visual contained a section labeled as the ``AI recommendation'' with an ``explanation'', participants expressed that they drew on multiple sources throughout the interface for the ``explanation'' of the recommendation. This included the ``explanation'' we provided but also included the patient's medical record and information from trusted clinical references, among other information provided in the interface---which they interpreted as a whole. Significantly, for participants, the implications from this information was important, not just for predicting medical consequences, but also for predicting legal consequences.

Participants' interpretation and integration of information across the AI system, their clinical experience, and other resources holds implications for the ways that we address users' legal considerations as we design XAI systems. Specifically, AI Creators should consider layering different kinds of information with algorithmically-created AI explanations to create more wholistic explanations. This is critical to adoption and building trust with physicians around the system's outputs. 

With many professionals, but especially physicians, different levels of trust correspond to different resources. For example, physicians place significant trust in FDA certifications on medical devices given the FDA's historic and rigorous evaluation processes. Physicians also significantly trust and rely on societal guidelines, which are based on scientific evidence to summarize the current medical knowledge, weigh the benefits and harms of diagnostic procedures and treatments, and give specific recommendations \cite{GuerraEtAl2023}. It is incorrect to assume that users---especially physicians who bear the ethical and legal consequences of medical decisions---will privilege an AI recommendation over other trusted sources. 

Consequently, AI Creators should re-frame their understanding of algorithmic explanations as one of many sources that physicians draw on. Considering how algorithmic outputs from an AI system align and are integrated with other trusted resources can significantly improve how AI explanations meet physicians' needs, helping physicians calibrate their trust in system outputs and predict the legal risks presented by the AI recommendation.

\paragraph{Surfacing Potential Blind Spots}
Our findings also show that physicians may not anticipate some legal and medical risks from AI systems that can dynamically evolve as it is used. Participants in our study anticipated risks largely based on their experiences with other physicians and their EHR system, which may point to potential blind spots as they anticipate errors that could harm patients. Participants referenced errors in current EHR systems when describing potential risks from an AI system basing its recommendation on potentially erroneous or messy information in the EHR. Similarly, errors such as data that lacks patient diversity or doesn't account for location-specific factors are also common EHR errors. Even mistakes around not ensuring the system meets their logic or standard of care reflects the general responsibility that physicians are trained to take over all aspects of patient care. 

While we described the AI system as one that would evolve over time, participants in our study did not anticipate risks that lawyers have with fully adaptive AI systems. For example, \citet{Sundholm2024} describes how current malpractice litigation may not cover all harms caused to a patient by a fully adaptive, opaque medical AI. He lays out a scenario in which there is an AI system that uses images of a patient's retina to detect a particular eye disease far better than most physicians. On the whole, the system continues to improve as it gathers data based on diagnoses. Unfortunately, while the overall performance increases, it eventually has worse performance on detecting one specific variation of retinal disease, and it misdiagnoses a patient. Other lawyers have noted similar concerns \cite{PriceAndCohen2023_locating, PriceEtAl2019, GerkeEtAl2020, Chan2021} to argue for reshaping malpractice litigation. This demonstrates the need to identify and address legal concerns---both that physicians are aware of and those of which they may not be aware.

\paragraph{Interacting Like Colleagues}
Participants indicated that they would treat the AI as they would treat other medical consulting physicians or guidelines. For example, V3 said, ``I think I'm used to leaning on things like UpToDate, which effectively could be made by AI. I don't even really check who wrote it or their credentials. So I would just think of it as like, a higher type of recommendation, and I don't think I would separate that AI versus UpToDate.'' V8 explained, ``It's not a matter of ever trusting an AI, but seeing whether it's logical and meets my standard of logic.'' These sentiments and those we described in the findings around how physicians thought of the AI as a ``thinking buddy'' mirror the relationship that physicians engage in when they consult another medical physician. The consultant can provide information and an opinion, but ultimately it is the initiating physicians' job to consider their patient's needs and make the final decisions around their care. 

Understanding how physicians may treat an AI system as they would another medical consultant, opens potential avenues for exploring the legal parallels between malpractice suits involving medical consultants. It can also help inform laws governing AI, which ultimately impacts physicians' use of AI systems. Understanding the perception of AI as a medical consultant can also inform how AI Creators develop AI explanations.  For example, it can help inform how AI Creators incorporate trusted resources and guide the kinds of information they should ensure they can output from explanation algorithms.

\section{Conclusions}

Health care professions are most often ultimately accountable for guaranteeing the performance of AI-based technologies, bearing the burden of decision making and the assumption of risk for harm \cite{MaddoxEtAl2025}.
While it's key to clearly communicate with physicians the liability risks around using an AI system, the implications for physician liability when using AI are unclear and debated. An article in the American Journal of Ophthalmology agreed with a 2019 statement by the American Medical Association, saying ``it is inappropriate for clinicians, using an autonomous AI to make a diagnosis they are not comfortable making themselves, to have full medical liability for harm caused by that AI... this paradigm shifts medical liability for a medical diagnostic from the provider...to the autonomous AI creator'' \cite{AbramoffEtAl2020}. This contrasts with conversations among legal scholars several of whom have argued for a more distributed model of risk among physicians, AI creators, admin, and other parties \cite{Sundholm2024, Price2022, Chan2021}. Given its potential impact on users, it's critical to identify legally salient information that users should know when using an AI system, even if the answer is that the risk is unknown or currently undetermined.

Through interviews with 10 physicians, our study identifies ways to understand and design AI explanations through a physician-centered perspective of their legal considerations. We describe how physicians are likely to fluidly consider information across the AI, XAI, interface, and the rest of their clinical environment to understand the medical and legal consequences of using an AI system or recommendation. While participants in our study highlighted the potential benefits and harms from AI systems, including legal and medical risks, they still failed to understand some legal risks, highlighting the spectrum of legal needs that may need to be addressed. Considering how physicians may be inclined to treat AI recommendations as they would treat other medical consultants opens pathways for exploring legal parallels and inform the creation of AI explanations, enabling physicians to protect patients and themselves.

\bibliography{aaai24}

\appendix

\newpage

\end{document}